\documentclass[draftcls,12pt,onecolumn,oneside]{IEEEtran}

\usepackage{graphicx}
\usepackage[numbers,sort&compress]{natbib}
\usepackage{booktabs}
\usepackage{multirow}
\usepackage{algorithm}
\usepackage{algorithmicx}
\usepackage{algpseudocode}
\usepackage{amsmath}
\usepackage{bm}
\usepackage{url}
\usepackage[justification=centering]{caption}
\usepackage{float}
\usepackage{subfigure}

\floatname{algorithm}{Algorithm}

\begin{document}

\title{Customized Slicing for 6G: Enforcing Artificial Intelligence on Resource Management}

\author{Wanqing Guan, Haijun Zhang, \IEEEmembership{Senior Member,~IEEE,} \\ and Victor C. M. Leung, \IEEEmembership{Fellow,~IEEE}
\thanks{Wanqing Guan and Haijun Zhang are with Beijing Engineering and Technology Research Center for Convergence Networks and Ubiquitous Services, Institute of Artificial Intelligence, University of Science and Technology Beijing, Beijing, China (E-mail: wanqingguan@ustb.edu.cn, haijunzhang@ieee.org). 

Wanqing Guan is also with State Key Laboratory of Media Convergence and Communication, Communication University of China, Beijing, China.

Victor C. M. Leung is with the Department of Electrical and Computer Engineering, The University of British Columbia, Vancouver, BC V6T 1Z4 Canada (E-mail: vleung@ece.ubc.ca).
}}

\maketitle

\begin{abstract}

Next generation wireless networks are expected to support diverse vertical industries and offer countless emerging use cases. To satisfy stringent requirements of diversified services, network slicing is developed, which enables service-oriented resource allocation by tailoring the infrastructure network into multiple logical networks. However, there are still some challenges in cross-domain multi-dimensional resource management for end-to-end (E2E) slices under the dynamic and uncertain environment. Trading off the revenue and cost of resource allocation while guaranteeing service quality is significant to tenants. Therefore, this article introduces a hierarchical resource management framework, utilizing deep reinforcement learning in admission control of resource requests from different tenants and resource adjustment within admitted slices for each tenant. Particularly, we first discuss the challenges in customized resource management of 6G. Second, the motivation and background are presented to explain why artificial intelligence (AI) is applied in resource customization of multi-tenant slicing. Third, E2E resource management is decomposed into two problems, multi-dimensional resource allocation decision based on slice-level feedback and real-time slice adaption aimed at avoiding service quality degradation. Simulation results demonstrate the effectiveness of AI-based customized slicing. Finally, several significant challenges that need to be addressed in practical implementation are investigated.

\end{abstract}

\section{Introduction}

In the coming era of 6G, the proliferation of new smart terminals and the explosion of new applications in vertical industry markets, such as augmented or virtual reality (AR/VR), unmanned aerial vehicles (UAV), fully autonomous driving, satellite-ground communications, etc., are forcing mobile network operators (MNOs) to carry the complex scenarios and deliver diverse services \cite{8412482}. Meanwhile, user demands are evolving continuously, making it more difficult to provide customized services and make personalized decisions in real time \cite{9040257}. 6G is expected to satisfy the dynamic and differentiated demands of users through real-time micro-management of multiple resources including communication, computing and storage resources. 

The concept of network slicing is proposed in 5G to create end-to-end (E2E) slice instances according to the different requirements of various services. As a key innovation expected to be inherited in 6G, network slicing is able to reduce the capital expenditure and operating expense (CAPEX/OPEX) by sharing the network resources among multiple tenants. Tenants, such as mobile virtual network operators (MVNOs), over-the-top (OTT) and vertical industries with limited capacity or coverage, rent the physical resources of MNOs or infrastructure network providers (InPs) to provide diversified services. To further reduce CAPEX/OPEX and increase revenue opportunities, tenants are motivated to unite the available resources provided by different InPs to enhance their attractiveness and acquire more subscribers \cite{7514161}. 

Therefore, there is a tremendous need to efficiently manage multi-dimensional resources of multi-InPs while meeting the strict and diversified service requirements of multi-tenant under dynamic environment. Creating customized slices for multiple tenants according to their preferences enables flexible and adaptive resource management \cite{8004168}. Moreover, allowing tenants to customize the resource allocation for each slice can dynamically adapt to the changes in network environment caused by user mobility, time-varying channel conditions and so on. However, supporting more 6G innovative services and satisfying increasingly-diverse user demands impose significant challenges for customized slicing, particularly in terms of E2E slices management and multi-dimensional resources orchestration. 

The first challenge is how to achieve real-time status observation of slices by depicting dynamic slice deployment and scalable resource utilization. Slice status information should be accurately obtained and quickly incorporated in decision-making of resource allocation. Then, efficient resources planning is conducted based on current status information of slices, including reserving resources for slices and determining the placement of virtual network functions (VNFs) for differentiated slices. 

Considering that an E2E slice consists of a number of interconnected VNFs from radio access network (RAN), core network and transport network, combinatorial optimization of numerous resources is the second challenge. The differences in profit of providing multiple resources to different tenants need to be accounted for when maximizing long-term revenue of InPs as network slice providers (NSPs). Striking a balance between the resources utilization of infrastructures and the profits of differentiated services provisioning is crucial for NSPs.

Last but not least, quickly satisfying the dynamic demands of differentiated services is another challenge. Since the scale and rates of network flows keep changing, the resources allocated to slices need to be adjusted in time to cope with the dynamic user demands. Additionally, the growing number and types of slices result in high complexity of slice adaption. Trading off the cost of reconfiguring slices and the satisfaction of stringent service quality becomes harder for tenants as network slice customers (NSCs).

Artificial intelligence (AI) saw rapid development during the past ten years and solved many pain points in different industries, such as healthcare, autonomous driving, smart manufacturing, etc. As one of the most promising AI tools, machine learning (ML) techniques have been widely applied in wireless communications \cite{8743390}. By iteratively learning from the reward feedback of environment, an optimal decision can be quickly achieved with ML methods compared to the conventional model-based optimization methods. Many researches adopt reinforcement learning (RL) based approach to manage resources involving both radio access part and core network part \cite{8540003}. RL incorporates farsighted system evolution into its decision-making and updates decision strategies to reach optimal performance through feedback of the previous decisions.

However, the existing RL based resource allocation methods \cite{8540003,8666109} and the AI-assisted network architecture for network slicing \cite{8954683} are inadequate to balance the capability of customizing resources for multiple tenants and maximizing revenues for multiple InPs. It is still very challenging to allocate resources across multiple domains and customize resources for each tenant simultaneously. In this article, we provide an AI-based hierarchical resource management framework, leveraging AI algorithms in both of the global management of multi-domain resources and the local slice adaption for multiple tenants. In addition, we introduce a customized slicing procedure for the proposed framework, realizing real-time on-demand resource provisioning and long-term revenue maximization. 

The remainder of this article is organized as follows. We first discuss the characteristics of customized slicing in the scenario of multi-InPs and multi-tenant, and explain why AI-based approaches is adopted. Then, a hierarchical framework is proposed for intelligent resource management of E2E slices, supporting slice customization for each tenant based on RL methods. The procedure of observing E2E slices' status and incorporating it into decision making as slice-level feedback is introduced. We illustrate the effectiveness of the prosposed intelligent management scheme in achieving high revenue and maintaining service quality. Finally, the challenges in practical implementation of 6G intelligent resource management are highlighted.

\section{Motivations and Background}

\subsection{Network Slicing Across Multiple Infrastructures}

As a fundamental attribute of 5G and beyond, network slicing is realized with the maturity of software defined networking (SDN) and network function virtualization (NFV). As the enablers of network slicing, NFV decouples software and hardware by virtualizing network functions and running them on the virtual machines (VMs) while SDN architecture provides centralized control plane for the configuration of network resources. These techniques prompt a service-based E2E wireless network architecture where VNFs of RANs and core network are placed as VMs deployed in data centers (DCs) of cloud InPs. The diverse demands of tenants can be satisfied through flexibly  managing resources and efficiently orchestrating VNFs of slices. By involving tenants in virtual network embedding (VNE) calculation, virtual networks could be provided in a tenant-driven manner with a trade-off between cost-effectiveness and time-efficiency. Since E2E slices require multiple resources, multiple domains administrated by different cloud InPs form a federated environment to jointly provide tenants with resources. 

E2E network slicing across multiple infrastructures has been discussed in the literature while management and orchestration (MANO) operations of slices in multiple administrative domains are also concern \cite{8758980}, as well as the life-cycle management operations. Through flexible slicing, heterogeneous resources of these cloud infrastructures can be utilized in a customized manner and the additional costs of the coalition can be reduced \cite{8088538}. Besides, analyzing the profit of resources provisioning and monitoring the status of resource utilization are essential in dynamic real-time E2E slicing. Specifically, performing admission control of resource requests needs to consider the revenue of NSPs as well as the service requirements and reallocating resources across multiple domains requires a global view of slice deployment status. To handle massive requests of configuring and modifying E2E slices dynamically, RL methods are used in our architecture, improving the speed and accuracy of decision-making.

\subsection{Multi-tenant Slicing}

Following the upcoming trends of applications, such as smart driving, AR/VR cloud gaming, the typical scenarios supported by 6G include further enhanced mobile broadband (FeMBB), ultra-massive machine-type communications (umMTC), extremely reliable and low-latency communications (eURLLC), long-distance and high-mobility communications (LDHMC), and extremely low-power communications (ELPC) \cite{8766143}. For services in these scenarios, guaranteeing extreme quality of experience (QoE) continuously requires rapid adjustment of network parameters based on real-time monitoring of network status. In order to guarantee QoE and boost revenue, efficient sharing of the underlaid network infrastructure has stimulated the interest of the research community \cite{9076108}. By establishing efficient network sharing schemes, multiple tenants which may own conflicting resource requirements obtain access to the different parts of the limited resources. 

As service providers, tenants rent resources to offer slice instances according to heterogeneous service requirements, which enhances the existent resource sharing flexibility. Network slicing allows various tenants to provide better-performing and cost-efficient services by supporting customized slices \cite{8540003}. Due to the uncertainty of service requirements, many model-free AI-based solutions are applied to jointly allocate multi-dimensional resources to slices \cite{8792072}. Moreover, RL has become an effective method to solve the decision-making problem of network slicing in the uncertain and probabilistic environment \cite{9057705}. For each tenant, when the traffic flow arrival/departure results in the degradation of slice’s service quality, individually deciding how to reallocate available resources is necessary. Owing that the traffic variation cannot be predicted without error, RL methods are also applied in slice adaption decisions of our architecture.

\subsection{Deep Reinforcement Learning}

Because resource allocation in wireless networks affects the QoE of services, various resource allocation methods have been studied over the past decades, including optimization, heuristic and game theoretic. As wireless networks become more complex, the static model-based algorithms will be inapplicable in the real dynamic network because of the long decision-making time and high computing burden. Owing to the capability of learning an optimal policy quickly, RL has been preferred for decision-making in the time-varying network environments and widely applied in solving many resource management problems, for instance, power control, spectrum management and computation resource management \cite{8743390}. As one of the most commonly adopted conventional RL algorithm, Q-learning suffers from slow convergence speed when the state space and action space are large. Deep reinforcement learning (DRL) algorithm which integrates deep neural network (DNN) with RL has been proposed by Google DeepMind, and the application of many advanced DRL algorithms has triggered tremendous research attention \cite{8931561}. 

\begin{figure}
  \centering
  \includegraphics[width=0.5\columnwidth]{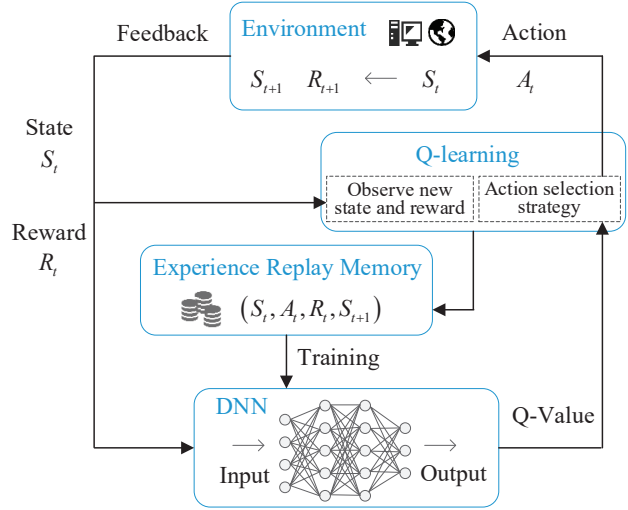}
  \caption{An illustration of deep Q-learning}
  \label{fig:DQN}
\end{figure}

Based on deep Q-network, DRL as shown in Fig. \ref{fig:DQN} outperforms conventional RL because experience replay is used to increase the efficiency of learning and enhance the stability of DNN. After performing action selection, reward calculation and new state observation, the mini-batches of experience are sampled uniformly at random to feed into the neural network during the learning process. DNN which is used to approximate the Q-value function takes the current states as the input and outputs a set of Q-values for all of the state-action pairs. Instead of using Q-table to store Q-values in the Q-learning algorithm, the deep convolutional network is used to address the instability caused by the correlations. Experience replay memory randomizes over the data, thereby allowing for greater efficiency and breaking the strong correlations between the samples. Hence, DNN improves the convergence of Q-learning and enables the deep Q-learning (DQL) algorithm to solve the problems which have a high-dimensional state-action space.

\section{AI-based Hierarchical Resource Management Framework}

\subsection{Hierarchical Resource Management}

\begin{figure}
  \centering
  \includegraphics[width=0.8\columnwidth]{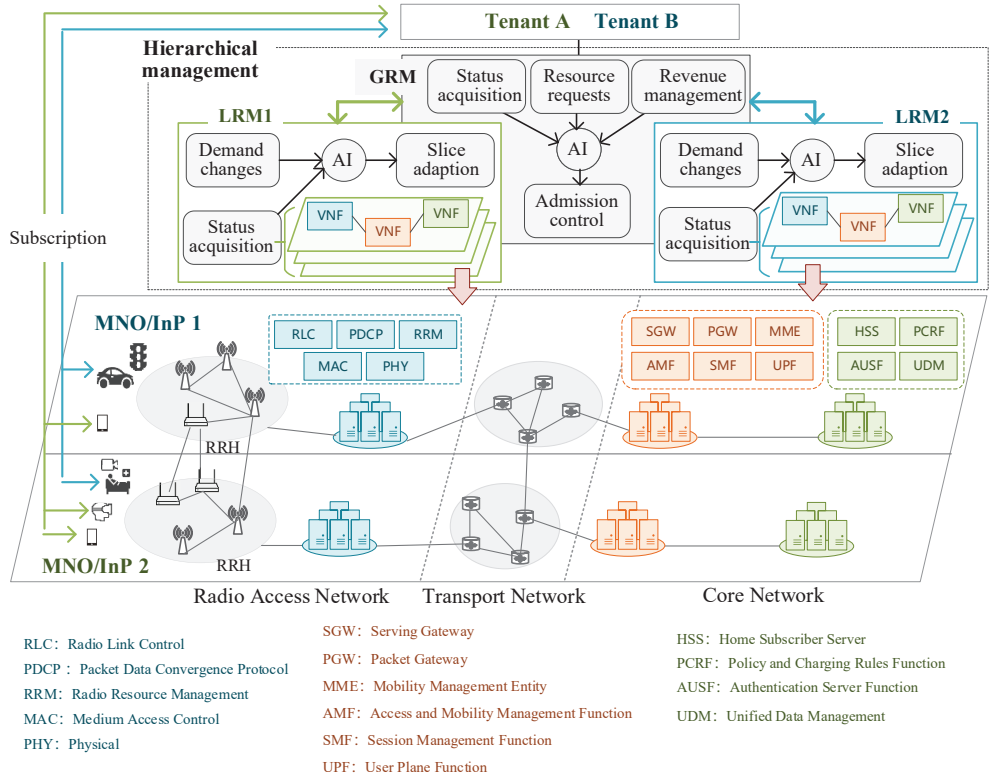}
  \caption{The AI-based hierarchical resource management framework}
  \label{fig:scenario}
\end{figure}

Planning deployment location from a global view can effectively avoid resource competition caused by the increasing number of co-located VMs on the same server. Scheduling multi-dimensional resources in a comprehensive and balanced way can potentially increase resource efficiency and avoid resource waste and shortage. In order to improve the revenue of resource providers, managing multi-domain resources centrally and allocating optimal amount resources to E2E slice instances are required. Given that traffic load variation of the slice might degrade QoE, the centralized management approach faces performance issues and limits the autonomy of the tenants. Hence, based on the MANO architecture for multi-domain slices in 5G \cite{8758980}, an AI-based hierarchical resource management framework shown in Fig. \ref{fig:scenario} is proposed to integrate intelligence in customized slicing for 6G use cases in the scenario of multi-InPs and multi-tenant.

To meet dynamically evolving service quality requirements and support fine-grained network decision optimization, the proposed framework introduces a global resource manager (GRM) to handle incoming differentiated resource requests from tenants, and multiple local resource manager (LRM) to deal with the demand changes in resource requirements for individual tenant. The deployment of GRM and LRMs enables two-layer customization of slices, which means that the resources are firstly allocated to each tenant according to the heterogeneous slice performance requirements, and then resource allocation to each slice is optimized and adjusted according to the real-time observation of demand changes. It is worth noting that the AI-based algorithms used in global resource allocation and local slice adaption can be different. 

The hierarchical approach can enable flexibility and scalability properties by distributing resource management to individual tenant. GRM maintains the overall control over the LRMs and delegates the concrete operations to each LRM. GRM is responsible for charging of slice owners and monitoring the LRMs while allocating federated resources across multiple domains. LRM performs slice adaption by adjusting the assigned resources to maintain service quality. Moreover, the LRMs not only provide each tenant the ability of resource customization, but also have the distinguishing feature of transmitting the status of slices to the GRM.

To handle with the traffic dynamics quickly, the status of slices which include the deployment location of VNFs and the condition of traffic flows passing through these VNFs are observed periodically. Mornitoring slice deployment and resource utilization facilitates to maintain service quality and enhance resource efficiency. Observing the real-time status of E2E slices provides a reference for determining whether or not to perform resource adaption. To realize real-time resource monitoring and slice topology information updating in the scenario of multi-InPs and multi-tenant, both of the differentiated slices provided by multiple tenants and the joint infrastructure network which consists of multiple infrastructures are depicted. Specifically, the cooperation between these infrastructure networks and the mapping relationships between the physical servers and the VNFs deployed in these servers are precisely delineated.

\subsection{Customized Slicing Procedure}

\begin{figure}
  \centering
  \includegraphics[width=0.6\columnwidth]{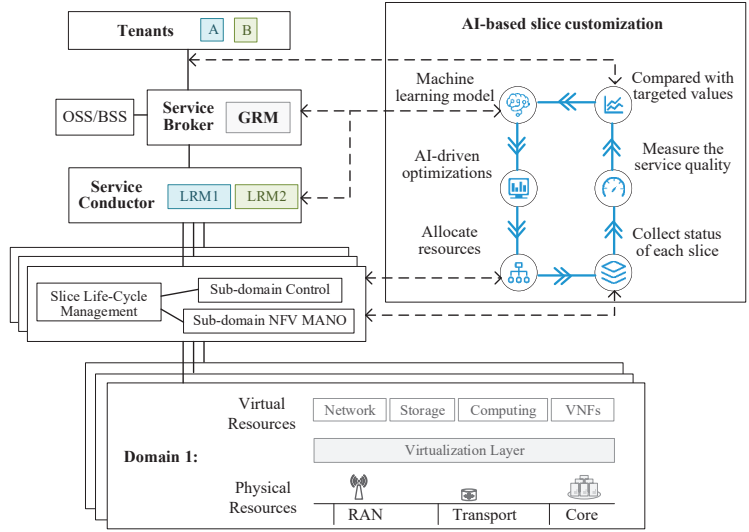}
  \caption{The procedure of customized slicing with the proposed framework}
  \label{fig:schedule}
\end{figure}

Figure \ref{fig:schedule} shows the procedure of customized slicing with the proposed AI-based management framework. After receiving the real-time slice requests from multiple tenants, the GRM deployed in the functional plane named the Service Broker performs admission control of these requests based on ML model. The NSPs make a trade off between the resource requirements associated to these requests and the revenue achieved by providing required resources. Multi-dimensional resources are allocated to tenants with the objective of maximizing the long-term revenue of NSPs. For the DRL-based resource allocation performed in GRM, states are defined as the number of accepted requests belonging to different tenants, actions taken by each agent are accepting/rejecting the arrival slice requests and reward is related to slice utility. With the output of the DRL algorithm in GRM, slices are deployed and the status of slices are recorded periodically.

Depending on perceived status, the current service quality can be measured and compared with its target quality requirements. The current service quality satisfaction reflects the gap with the target value. The target values regarded as the desired service quality should be defined as the level of service that the available resources of InPs can and should provide, thus they are preset and fed as input to the optimization problem of slice adaption. As the slice-level feedback, current service quality satisfaction is used to improve the performance of ML model and update the model with demand changes that could occur over time. When there are changes in slice requests, such as a sudden increase in resource requirements, the ML model is utilized to maintain service quality for admitted slices by micro-managing resources.

The motivation of performing resource adaption generates from the mismatch between available resources and the varying traffic demand in the slice. This mismatch might cause two kinds of issues, one is that available resources are exhausted and the other is that partial resources are idle. The former means unfair resource allocation resulting in the low data rate of newly accepted user, and the latter means that the revenue of tenant is declining. To avoid unbalanced distribution of available resources, i.e., some resources are under-utilized, some are over-utilized, the allocated resource of each tenant should be adjusted to maximize the profits of available resources. After receiving the requests of adjusting resources for multiple slices, tenant makes decisions by weighing the cost and revenue of adjusting resources for each slice to ensure optimal resource efficiency.

The LRM deployed in Service Conductor performs the DRL-based slice adaption. States are tied to the current service quality satisfaction and actions denotes whether slice adaption is permitted. Reward is defined as the revenue obtained by adjusting resource minus the resource consumption cost and operational cost. The revenue is related to the amount of money paid by the service subscribers for guaranteeing service quality, which depends on the type of slice.  The resource consumption cost represents the cost of providing more resources, such as the extra processing units required by the newly arrived traffic flows. The operational cost means the cost of performing reconfiguration, which includes the cost of service interruption caused by reallocating resources and migrating VNFs among physical servers. There is no doubt that DQL used in this article can be replaced by other advanced DQL-based algorithms to achieve better performance.

\section{Evaluation}

Having introduced the key elements of the proposed framework, the next important step is to evaluate the performance of customized slicing and verify the benefits of AI-based resource management. In this section, the effectiveness of the proposed management framework is validated in terms of improving the long-term revenue and guaranteeing service quality.

\subsection{Experimental Setup}

The AI-based resource management approach is implemented in python where the Tensorflow library is used to build the ML model. For the purpose of comparison, a non-intelligent resource management approach with centralized resource allocation of differentiated slices is also implemented. DQL algorithm is compared with a greedy algorithm used in the non-intelligent resource management framework. The greedy algorithm permits resource reallocation so long as the remaining resources are enough, which ignores difference in the value of reconfiguring differentiated slice. The AI-based framework reserves more resources for slices which can bring more revenue, and performs slice adaption in a cost-effective manner. To verify the performance of the proposed framework, there are two tenants which owns different types of slices in the simulation, and the flow dynamics of these two types of slices are different. Assuming that slice of type 1 owned by tenant A and slice of type 2 owned by tenant B have the same total number of VNFs, and the VNFs are deployed in DCs of the joint infrastructure network provided by two InPs at the beginning of simulation.

\begin{table}
  \centering
  \caption{Simulation parameters}
  \label{tab:1}
    \begin{tabular}{ccccc}
    \toprule
    Items & Values \\
    \midrule
    Total number of physical nodes & $10$ \\
    Number of DCs & $5$ \\
    Resources of each DC & $300$ processing units \\
    Number of VNFs for each slice & $4$  \\
    Number of flows for slice of type 1 & $180$  \\
    Flow arrival interval for slice of type 1 & $2$ (sec) \\
    Flow service time for slice of type 1 & $200$ (sec) \\
    Number of flows for slice of type 2 & $60$  \\
    Flow arrival interval for slice of type 2 & $5$ (sec) \\
    Flow service time for slice of type 2 & $300$ (sec) \\
    \bottomrule
    \end{tabular}
\end{table}

A summary of the simulation parameters is listed in Table 1. The topology of each infrastructure network is generated according to the algorithm of Barabási-Albert (BA) scale-free networks because a forthcoming node of the communication network tends to connect itself to the nodes with large degrees. While resources in wireless networks are miscellaneous, here we confine resources to computational resource of DCs. We assume that processing one unit of data flow requires one unit of computational capacity. There are 5 DCs in the joint infrastructure network with 10-node topology and each DC has capacities of 300 processing units. The flows in each slice arrive following a Poisson process, and the service time follows an exponential distribution (the arrival and departure rates are given in Table \ref{tab:1}). The status of slices are recorded and stored in a database. Service quality satisfaction of slices are measured periodically and the period between two measurements is 1s. In this section, to calculate the values of current satisfaction, the number of waiting data flows in each moment of measurement is recorded and normalized.

\subsection{Long-term Revenue}

First, the proposed framework is able to maximize the average reward by reserving resource for resource requests which could bring more revenue. According to the requests of tenants, GRM achieves the optimal decision-making for admission control through DRL-based algorithm. The arrival rates of resource requests from tenant A and tenant B are set at 10 requests/hour and 12 requests/hour respectively while the completion rates of requests are set at 6 requests/hour. The immediate reward obtained by accepting resource request from tenant A is set at 2 and tenant B is varied while each resource request requires 60 processing units of each DC. 

\begin{figure}
  \centering
  \subfigure[Long-term revenue]{
    \includegraphics[width=0.4\columnwidth]{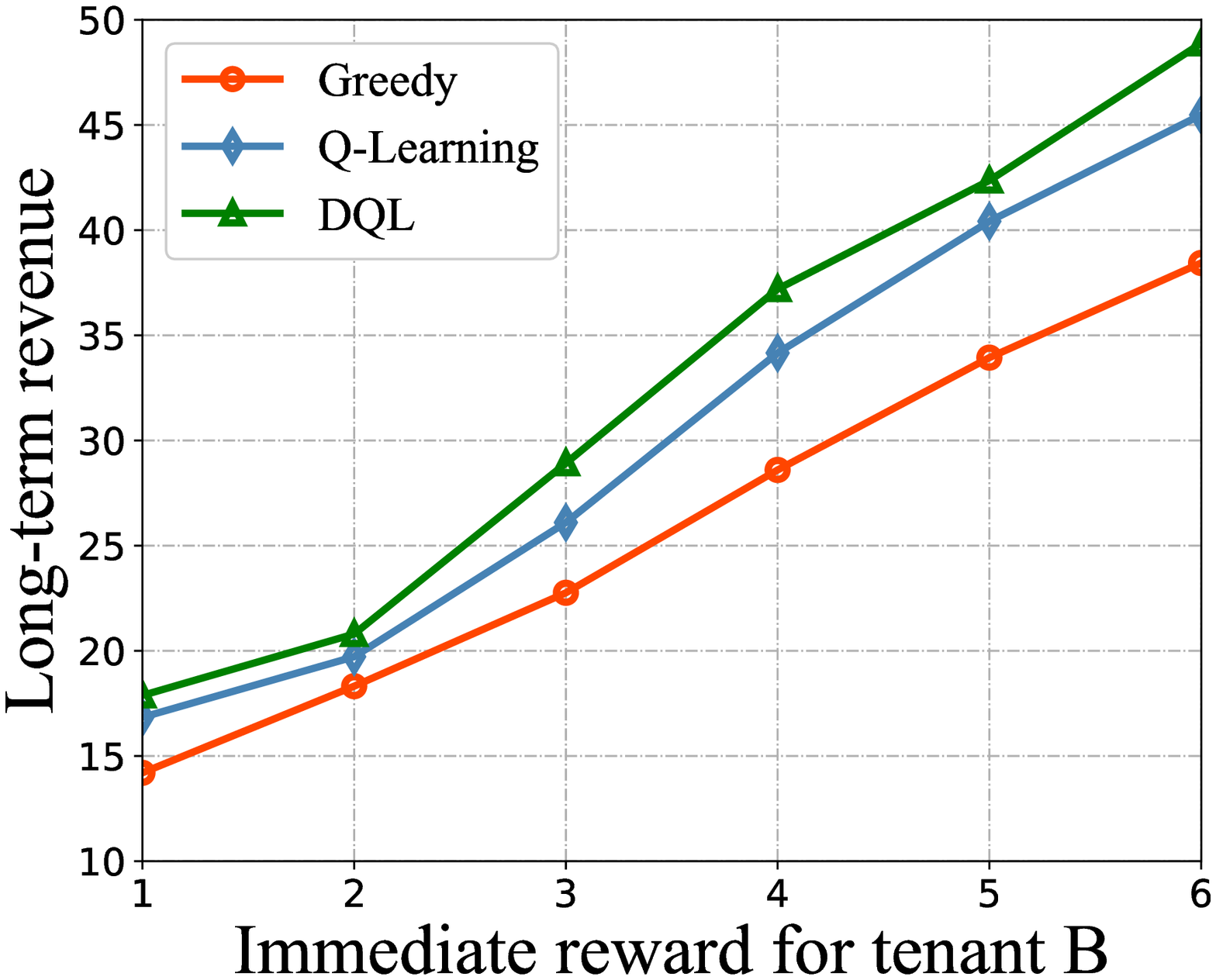}
    \label{fig:ltr}
  }
  \subfigure[Proportion of accepted requests]{
    \includegraphics[width=0.4\columnwidth]{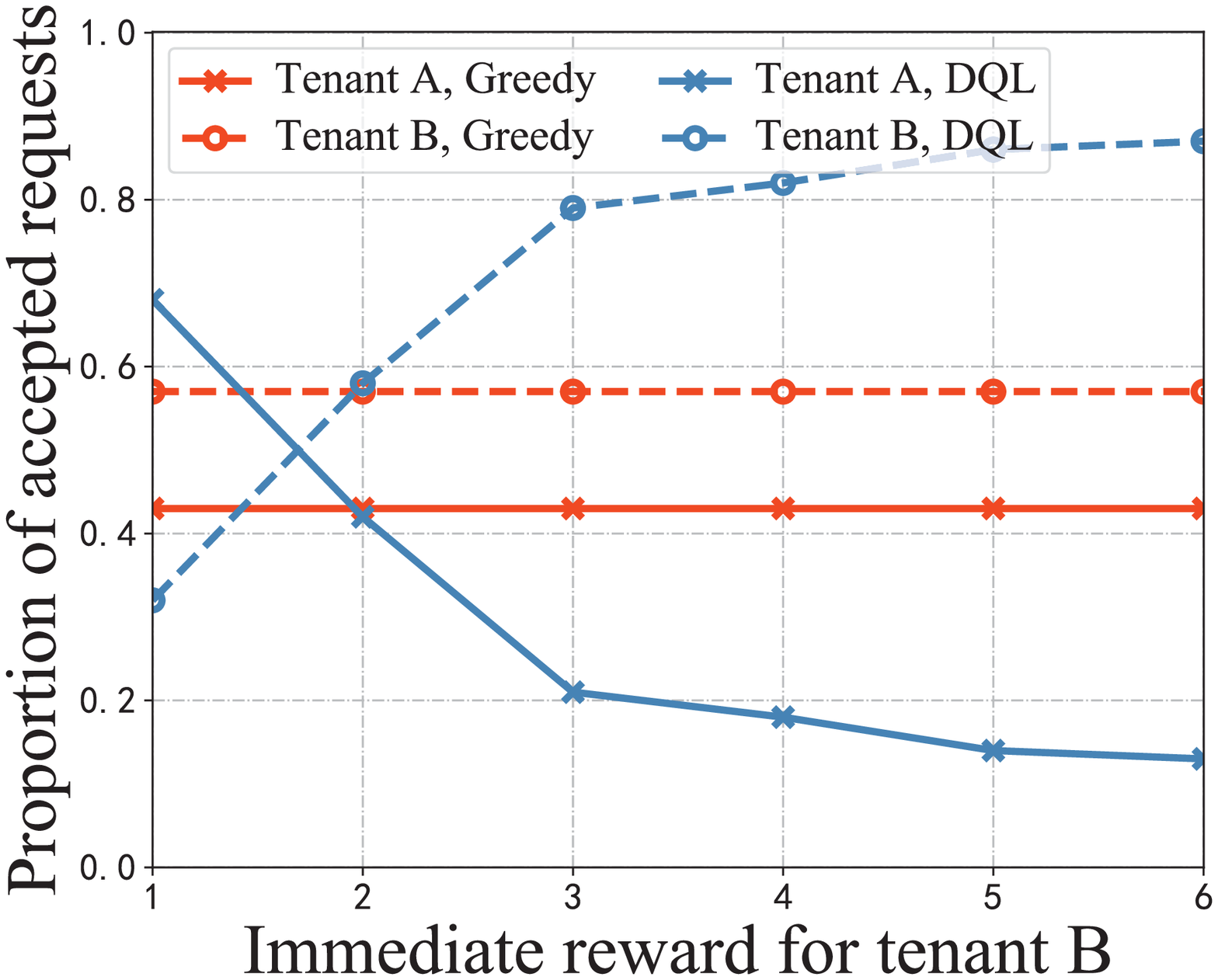}
    \label{fig:par}
  }
  \caption{The performance of the proposed framework when the immediate reward is varied.}
  \label{fig:dif_para}
\end{figure}

In Fig. \ref{fig:ltr}, the performance of the DQL algorithm with Q-learning and greedy algorithms in terms of long-term revenue are compared. It is shown that the revenue obtained by three algorithms are increased when the immediate reward of accepting requests from tenant B is varied from 1 to 6. However, the revenue obtained by RL algorithms, i.e., DQL and Q-learning, is significantly higher than that of the greedy algorithm. The reason is that RL algorithms reserve resource for the requests which may bring high reward while the greedy algorithm accepts the requests according to the available resource without considering reward. Besides, due to the slow convergence rate of Q-learning, the DQL algorithm achieves higher revenue than Q-learning.

To further analyze the performance of the proposed framework, the proportion of accepted requests from two tenants are calculated and shown in Fig. \ref{fig:par}. It can be observed that RL algorithms are likely to accept the resource requests which have higher immediate reward. For example, when the immediate reward of accepting requests from tenant B is larger than that from tenant A, there are more accepted requests from tenant B. In contrast, the greedy algorithm accepts request when the available resource satisfy the demand. Hence, the composition ratio of accepted requests is not affected by the change of the immediate reward.

\subsection{Service Quality Satisfaction}

Second, to substantiate that the proposed framework is able to maintain service quality, the service quality satisfaction of the intelligent framework and non-intelligent framework are compared. The satisfaction related to E2E delay is influenced by the remaining processing units allocated to this slice, which determines whether the incoming data flows can be delivered in time. Once the remaining resources are insufficient to meet the service quality requirements, the number of the waiting data flows is going to accumulate. When the allocated resource can not satisfy the dynamic demand, the slice adaption algorithm in each LRM of the proposed framework will be triggered to make decisions of reallocating processing units. With resource customization, service quality of slice could be maintained in a higher level by sacrificing the signaling cost and the communication overhead.

\begin{figure}
  \centering
  \subfigure[Service quality satisfaction]{
    \includegraphics[width=0.4\columnwidth]{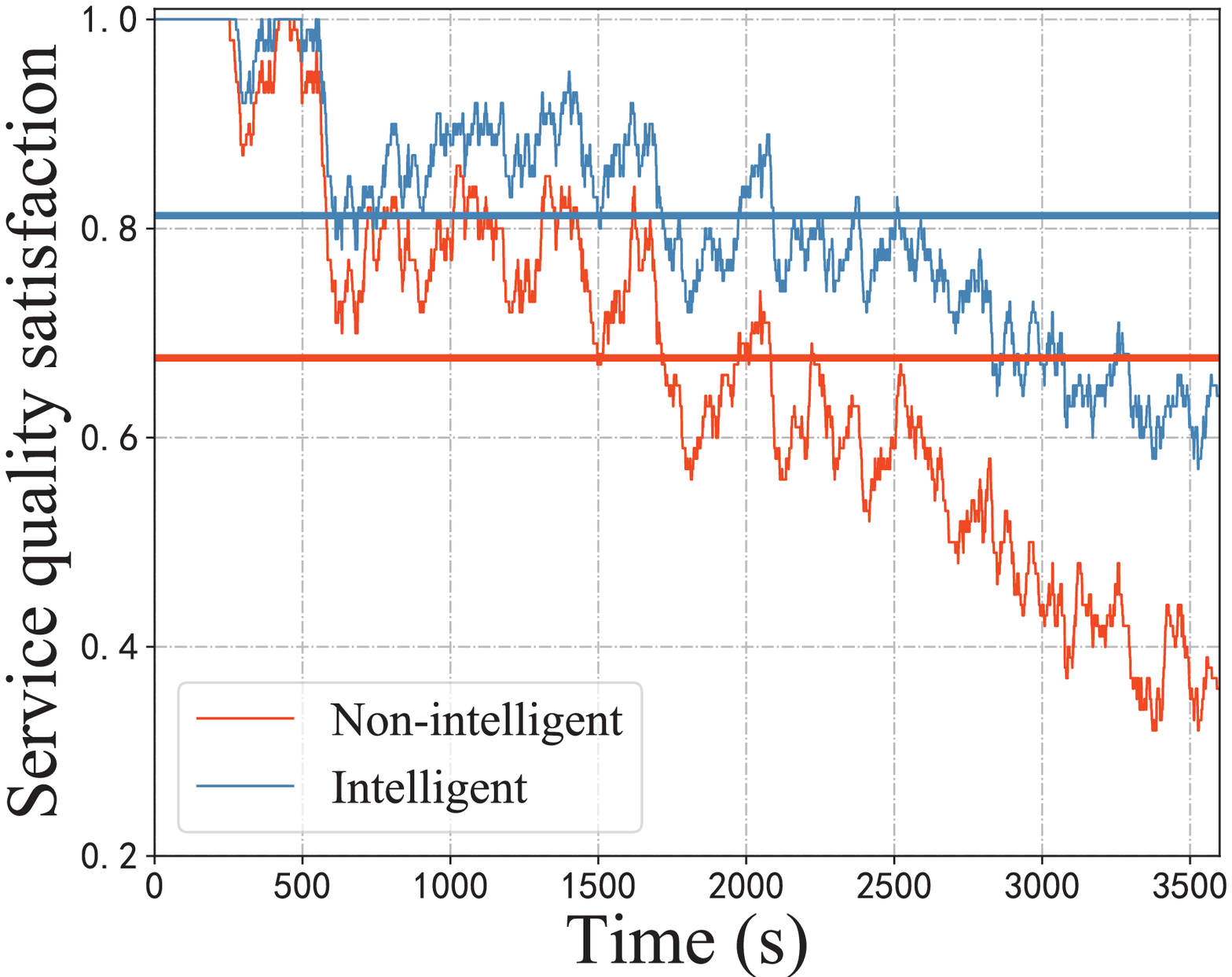}
    \label{fig:sa}
  }
  \subfigure[Decision delay]{
    \includegraphics[width=0.4\columnwidth]{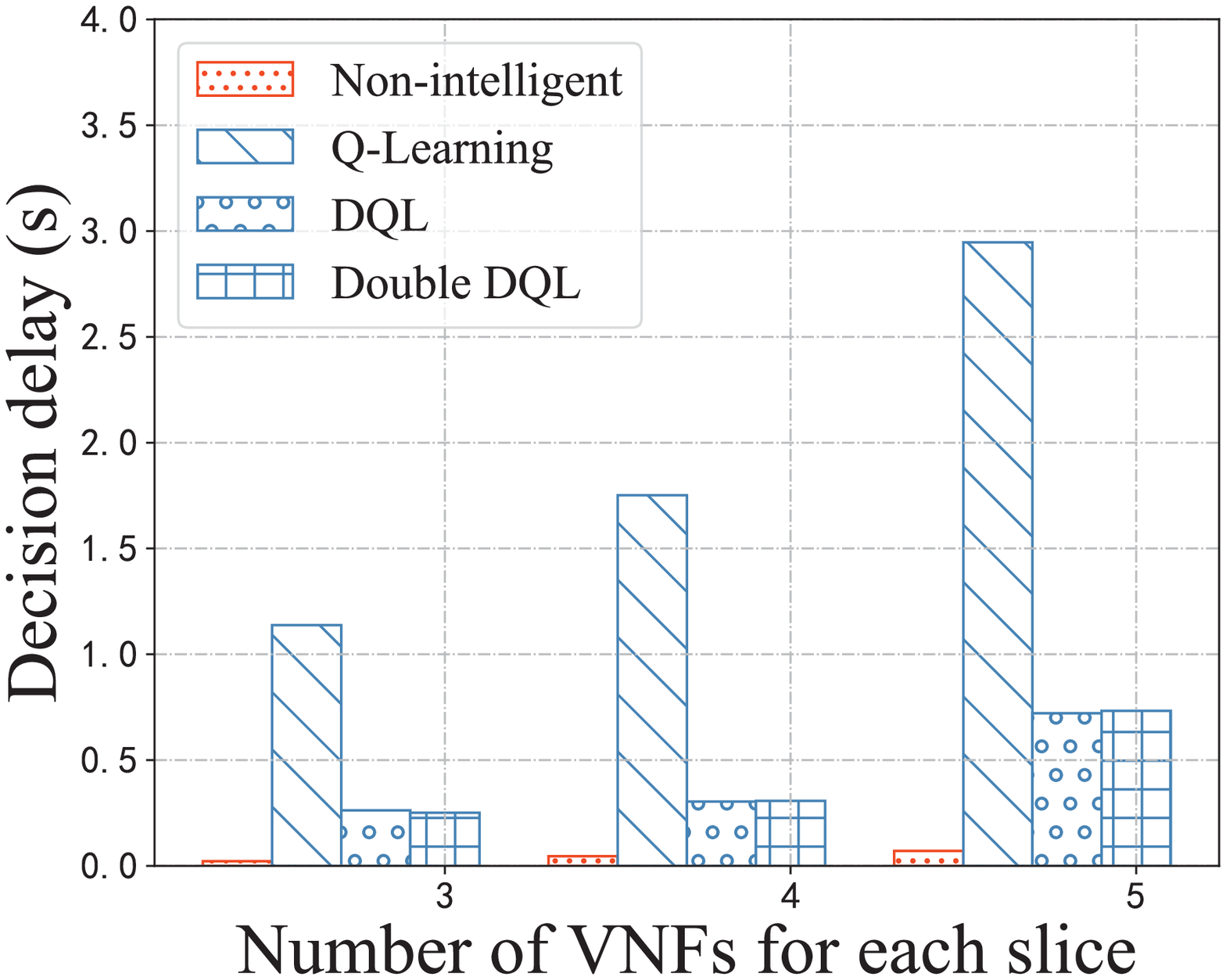}
    \label{fig:de}
  }
  \caption{Comparison between the intelligent and non-intelligent frameworks. (a) Service quality satisfaction vs. time; (b) Decision delay vs. the number of VNFs for each slice.}
  \label{fig:comp}
\end{figure}

In Fig. \ref{fig:comp}, we plot the service quality satisfaction vs. time and decision delay vs. the number of VNFs for each slice for the intelligent resource management framework and the non-intelligent framework. As shown in Fig. \ref{fig:sa}, the average satisfaction of the intelligent framework is increased by nearly 20\% with the proposed method, compared with the non-intelligent framework. In Fig. \ref{fig:de}, it can be seen that the non-intelligent framework requires minimum decision delay while the proposed framework with Q-learning algorithm requires the most computation time. In particular, more decision time is required as the number of VNFs increases, because reconfiguring slices becomes more complex. However, both DQL and double DQL which can further improve stability and avoid overestimation have decision delays of less than 1 second.

\section{Open Issues and Future Challenges}

With the deployment of GRM and LRMs, the proposed framework achieves multi-tenant oriented intelligent resource management with the aim of optimizing the long-term revenue of the NSPs, and realizes fine-grained resource customization with the aim of maintaining the service quality of slices from different tenants. In addition, AI-enabled technique or, more precisely, the use of ML algorithms allow the resource management framework to adapt the changes in resource requirements and learn optimal policy from the dynamic environment. Nevertheless, there are plenty of open issues that need further study and several challenges in practical implementation. Some of these issues and challenges are introduced below.

\emph{Real-time prediction of evolving user demands}: User demands are highly dynamic and uncertain. Therefore, there are efforts in the literature to predict users' behavior. However, correlating the evolutional tendency of demands to resource allocation in network slicing constitutes a challenging but interesting line of research.

\emph{Fast implementation of E2E slicing}: Responding to user demands in a real-time manner is essential in providing better service quality. For this reason, E2E slices need to be instantiated rapidly and completely. Therefore, more research should be conducted to develop practical solutions supporting easier experimentation in the scenario of multi-InPs and multi-tenant.

\emph{Adaptive adjustment of AI-based solution}: In the learning stage of AI-based resource management framework, the relatively long convergence time of ML methods undermines their usefulness. Besides convergence, the stochastic nature of the wireless network may require ongoing updates of the parameters and continuous adaption of ML methods. Therefore, feasible and scalable ML algorithms need more study and analysis.
 
\emph{Coordinated collaboration of multiple InPs}: As mentioned earlier, the cooperation between different infrastructure networks offers an attractive mean of providing multiple resources at low cost. However, the process of building effective collaboration is extremely challenging, which necessitates more investigation about management interfaces and resource isolation across multiple administrative domains.

\section{Conclusion}

This article has proposed intelligent resource management framework to enable customized slicing in the scenario of multi-InPs and multi-tenant based on the slice-level feedback and real-time service quality. Along with artificial intelligence, we propose a hierarchical framework where global resource manager utilizes machine-learning-based method coupled with service quality evaluation to make optimal resource allocation decisions and local resource managers collect status information about slice deployment and resource utilization to support fine-grained resource adaption. The simulation results show the effectiveness of the proposed reinforcement-learning-based resource management approach in terms of achieving high revenue of network slice providers and maintaining high quality of end-to-end services.

\ifCLASSOPTIONcaptionsoff
  \newpage
\fi

\footnotesize

\bibliographystyle{IEEEtran}

\bibliography{IEEEabrv,bibl}

\end{document}